\AtBeginDocument{%
  \paperwidth=\dimexpr
    1in + \oddsidemargin
    + \textwidth
    + 1in + \oddsidemargin
  \relax
  \paperheight=\dimexpr
    1in + \topmargin
    + \headheight + \headsep
    + \textheight
    + 1in + \topmargin
  \relax
  \usepackage[pass]{geometry}\relax
}
\RequirePackage{fix-cm}
\documentclass{svjour3}                     
\usepackage{graphicx}
\usepackage{amsmath}
\usepackage{subfigure}
\usepackage{float}
\usepackage{tensor}
\usepackage{soul}

\newcommand{\sgn}{\textrm{sgn}}

\begin{document}

\title{Geometric and physical properties of closed ever expanding dust models 
}


\author{Sebasti\'an N\'ajera         \and
        Roberto A. Sussman 
}


\institute{Sebasti\'an N\'ajera \at
              Instituto de Ciencias Nucleares, Universidad Nacional Aut\'onoma de M\'exico
(ICN--UNAM),\\ A. P. 70 –- 543, 04510 M\'exico D. F., M\'exico. \\
              \email{sebastian.najera@correo.nucleares.unam.mx} 
}

\date{Received: date / Accepted: date}

\maketitle

\begin{abstract}
Current observations suggest that our Universe is not incompatible with a small positive spatial curvature that can be associated with rest frames having a ``closed'' standard topology. We examine a toy model generalisation of the $\Lambda$CDM model in the form of ever expanding Lema\^{i}tre-Tolman-Bondi (LTB) models with positive spatial curvature. It is well known that such models with $\Lambda=0$ exhibit a thin layer distribution at the turning values of the area distance that must be studied through the Israel-Lanczos formalism.  We find that this distributional source exhibits an unphysical behaviour for large cosmic times and its presence can be detected observationally. However, these unphysical features can always be avoided by assuming $\Lambda >0$. While these LTB models are very simplified, we believe that these results provide a simple argument favouring the assumption of a nonzero positive cosmological constant in cosmological models.
\end{abstract}

\section{Introduction}
\label{intro}

The spherically symmetric exact solutions of Einstein's equations known as the Lema\^{i}tre-Tolman-Bondi (LTB) dust models are useful \emph{toy models} to study observational issues and structure formation in a Friedman Lema\^\i tre Robertson Walker (FLRW) background. If we assume $\Lambda>0$ these models provide a simple inhomogeneous generalisation of the  $\Lambda$CDM model favoured by current observations. In fact, models with $\Lambda=0$ and $\Lambda >0$ provide simple descriptions of a single CDM structure (overdensity or density void) in an FLRW background. The evolution of such structures can always be mapped rigorously to the formalism of gauge invariant cosmological perturbations (see comprehensive discussion in \cite{Sussman3,Sussman4}). As shown in \cite{Sussman3,Sussman4} (see also \cite{Sussman1,Sussman2}), LTB inhomogeneities can be described as covariant exact fluctuations that in their linear regime reduce to linear cosmological perturbations in the isochronous comoving gauge.        

Models with $\Lambda=0$ and $\Lambda >0$ provide also provide  simple relativistic generalisations of the Newtonian \emph{spherical collapse model}, which provide order of magnitude estimations of collapsing times and density contrasts that are useful in the design of numerical N--body simulations. See discussion and examples in \cite{Bolejko,Padm,Peebles}.

Ever expanding FLRW models with a closed topology (rest frames difeomorphic to the 3--sphere $\bf{S}^3$) and a dust source are not possible unless we assume that $\Lambda>0$. If $\Lambda=0$ then all closed FLRW dust models must have positive spatial curvature and must bounce and re--collapse. However, for LTB models the extra degrees of freedom decouple kinematic evolution and the topology of the rest frames, allowing (in principle) for ever expanding closed models even if $\Lambda=0$. In the 1980's when a nonzero  cosmological constant was not favoured, Bonnor \cite{Bonnor} showed interest in looking at ever expanding LTB models with $\Lambda=0$ and a closed topology. He showed that these models exhibit a thin layer surface matter distribution at a timelike hypersurface marked by the \emph{turning value} of the area radius (the ``equator'' of $\bf{S}^3$). Using the Israel-Lanczos formalism,  Bonnor derived the equation of state for this surface layer matter--energy distribution, regarding it in a pointblank manner as unphysical because it involved negative surface pressure (these were the times before dark energy). Hence, Bonnor concluded that full regularity of closed LTB models with $\Lambda=0$ required re-collapse and thus excluded ever expanding kinematics. More recent research allows for the interpretation of the negative surface layer pressure as surface tension \cite{Schmidt}.

In the present article we extend Bonnor's work by (i) showing that fully regular closed and ever expanding LTB models are possible once we consider $\Lambda>0$ and (ii) by looking for the case $\Lambda=0$ at the time evolution of the distributional surface source in comparison with the evolution of the continuous density. We show for models with zero and negative spatial curvature that the behaviour of this source is unphysical, since for large times the continuous dust density surface density decays at a much faster rate than the distributional surface density (which has no contribution to the quasilocal mass integral). In particular, we show that the presence of such distributional source would be detectable by observations through the redshift from sources connected by radial null geodesics that cross the equatorial hypersurface of $\bf{S}^3$. While the redshift as a function of comoving radius is continuous, its derivative is not, with the abrupt change of rate occurring precisely at this hypersurface. We show that this effect does not occur for re-collapsing LTB models with closed topology (for which there is no distributional source at the equator of $\bf{S}^3$). 

Since observations do not rule out a Universe whose rest frames have a closed $\bf{S}^3$ topology associated with a very small positive spatial curvature, then an LTB model with $\Lambda>0$ is a viable toy model approximation to a $\Lambda$CDM model that is favoured by observations. Hence, we argue that the results of the present article provide another argument to support the need for a positive cosmological constant, since without the latter all ever expanding CDM dominated models would be incompatible with a closed $\bf{S}^3$ topology.                

The section by section description of the paper is as follows. In section \ref{sec:LTBLam0} we provide a brief introduction to generic LTB models with $\Lambda=0$, while in section \ref{sec:closed} we examine the specific case of closed models. In section \ref{sec:regularity} we review Bonnor's work, section \ref{sec:surften} briefly introduces surface tension in curved spacetime and in section \ref{sec:lanczos} we provide an example of ever expanding closed models with zero and negative spatial curvature. The surface layer density is evaluated for these models, showing that in the large time regime the continuous density decays much faster than the surface density, which is an unphysical behaviour. We show in section \ref{sec:lamg0} that fully regular ever expanding closed models with $\Lambda >0$ are always possible. In sections \ref{sec:geod} and \ref{sec:redshift} we compute null radial geodesics for the spatially flat case in order to examine the observational detection of the thin shell distribution. Finally, in section \ref{sec:Kpos}, we show that no observational effects occur in the the case of re--collapsing models with positive spatial curvature and $\Lambda=0$, for which no thin shell distributional source arise.

\section{LTB Models with $\Lambda=0$}\label{sec:LTBLam0}

LTB models are exact spherically symmetric solutions of Einstein's equations with an inhomogeneous dust source with or without cosmological constant\footnote{We examine the case $\Lambda>0$ in section \ref{sec:lamg0}. Everywhere else, unless specifically stated, we assume $\Lambda=0$.}. This solutions are described by the LTB metric in comoving coordinates 
\begin{equation}\label{metLTB}
ds^2=-dt^2+\frac{R'^2}{1-K}dr^2+R^2d\Omega^2,
\end{equation}
where $R=R(t,r)$, $K=K(r)$ and $R'=\partial R/\partial r$. The field equations yield:
\begin{equation}
\dot{R}^2=\frac{2M}{R}-K, \;\;\;\;\; 2M'=8\pi\rho R^2 R',\label{ypunto}\\
\end{equation}
where $M=M(r)$ is the Misner--Sharp quasi--local mass--energy function, a well known invariant in a spherically symmetric spacetime, and $\dot{R}=\tensor{u}{^a}\tensor{\nabla}{_a}R=\partial R/\partial t$. The first equation in \eqref{ypunto}, a Friedman--like evolution equation, leads to a classification of the models in three kinematic classes according to the sign of $K=K(r)$, which determines the existence of a zero of $\dot{R}^2$, and thus, the kinematic evolution: for $K>0$ the models expand initially $\dot R>0$, reach a maximal expansion value $R_{\max}=2M/K$ where $\dot R=0$ and then collapse $\dot R<0$, while for $K\leq 0$ the models are ever expanding. Since $K=K(r)$, it is possible to have in a single model regions with different kinematic class (see comprehensive discussion in \cite{Humphreys}). 

The solutions of the Friedman--like equation in  \eqref{ypunto} define the kinematic classes as elliptic ($K>0$), hyperbolic ($K<0$) and parabolic ($K=0$) solutions given by
\begin{eqnarray}
K> 0:&{}& \;\;\;\; R=\frac{M}{K}(1-\cos \eta),\;\;\;\; \eta-\sin \eta =\frac{K^\frac{3}{2}}{M}(t-t_{bb}(r)),\label{ellipclass}
\\
K<0: &{}&\;\;\;\; R=\frac{M}{\vert K\vert}(\cosh \eta-1),\;\;\;\; \eta-\sinh \eta =\frac{|K|^\frac{3}{2}}{M}(t-t_{bb}(r)),\label{hyperclass}
\\
K=0:&{}& \;\;\;\; R=\left[\frac{9}{2}M(t-t_{bb}(r))^2\right]^\frac{1}{3},\label{paraclass}
\end{eqnarray}
with $t_{bb}(r)$ denoting the Big Bang time function such that $R(t_{bb}(r),r)=0$ for variable $r$ (notice that in general $t_{bb}'\neq 0 $). 

To fully determine an LTB model we need to prescribe the three free functions $M(r)$, $K(r)$ and $t_{bb}(r)$. Since the metric is invariant under rescalings of $r$, it is always possible to reduce this set of free functions to a pair of independent irreducible free functions. Given a choice of free functions, all relevant quantities of the models can be computed from the solutions for \eqref{ellipclass} to \eqref{paraclass}.

\section{Ever expanding closed models} \label{sec:closed}

Closed LTB models are characterised by rest frames that are compact 3--dimensional submanifolds without a boundary and with finite proper volume, which implies two possibilities: the rest frames are diffeomorphic to $S^3$ or to a 3-torus (an example of how to select the free functions latter case is given in \cite{Humphreys}). Since LTB models are spherically symmetric, the  topological class of the rest frames is directly connected with the existence of symmetry centers, which are regular timelike comoving worldlines $r=r_c$ generated by the fixed points of \emph{SO}(3), and thus comply with
\begin{equation}\label{condsym}
R(t,r_c)=\dot{R}(t,r_c)=0.
\end{equation}
Closed models diffeomorphic to ${\bf S}^3$ admit two symmetry centers, while rest frames with toroidal topology admit no symmetry centers. In closed LTB models the condition (\ref{condsym}) holds for two values of $r$, which can be denoted by $r=0$ and $r=r_c$. Since ${\bf S}^3$ is smooth there must exist a turning value $r=r_*$ such that $R'(t,r_*)=0$. Regularity conditions implies that $M=K=0$ and all radial gradients vanish at both symmetry centers.

\subsection{Regularity of ever expanding closed models} \label{sec:regularity}
 
After looking at closed LTB models with zero cosmological constant, Bonnor \cite{Bonnor} concluded that all ``physically acceptable closed models'' (PACM) must be elliptic everywhere and eventually, collapse. Bonnor defined a PACM by the following conditions:
\begin{enumerate}
\item $\rho$ is finite and non--negative
\item There are no comoving surface layers nor shell--crossing singularities.\label{condi2}
\item $K$, $M$ and $R$ are $C^1$.
\item $K$ satisfies extra regularity conditions at the symmetry centers, see \cite{Bonnor}.
\end{enumerate}
These conditions imply 
\begin{equation}\label{condireg}
\sgn(R')=\sgn(M')=\sgn(K'),
\end{equation}
and, as an immediate consequence, if zeroes of $R', M', \sqrt{1-K}$ exist, they must all be common and of the same order. If the zeroes of $R'$ are  different from the zeros of the other quantities, then shell crossings occur where the density and curvature scalars diverge with $R>0$. The second equation in \eqref{ypunto} together with \eqref{condireg} imply that the density is non--negative and bounded everywhere, except at the coordinate locus of a central singularity. The necessary and sufficient conditions to avoid shell--crossing singularities, as required by \eqref{condireg}, are given by the Hellaby--Lake conditions given explicitly in  \cite{Sussman2,Hellaby}.


\section{Lanczos--Israel formalism for closed models}\label{sec:lanczos}

In what follows we use Taub's approach to tensorial distributions within the  Lanczos-Israel-formalism \cite{Taub}. As proven in \cite{Senovilla1},  Bianchi's second identity holds in the distributional sense, therefore the following conservation equation holds in the sense of distributions:\, $\tensor{\nabla}{^a}\tensor{\underline{G}}{_a_b}=0$, where $\tensor{\underline{G}}{_a_b}$ is the distributional Einstein tensor. Following \cite{Senovilla2}, we discuss the relations that emerge from this conservation equation. 

Let $(\Sigma, h)$ be a hypersurface embedded in the $4$-dimensional spacetime $(M,g)$, the following equations are satisfied
\begin{eqnarray}
\tensor{\mathcal{G}}{_a_b}&=&-[\tensor{K}{_a_b}]+\tensor{h}{_a_b}[\tensor{K}{^c_c}],\label{distribfe}\\
\tensor{n}{^b}\tensor{\mathcal{G}}{_a_b}&=&0,\\
(\tensor{K}{^+_a_b}+\tensor{K}{^-_a_b})\tensor{\mathcal{G}}{^a^b}&=&2\tensor{n}{^c}\tensor{n}{^d}[\tensor{\underline{G}}{_c_d}]\label{eqsenov1}\\
\tensor{\overline{\nabla}}{^a}\tensor{\mathcal{G}}{_a_b}&=&-\tensor{n}{^c}\tensor{h}{^d_b}[\tensor{\underline{G}}{_c_d}],\label{eqsenov2}
\end{eqnarray}
where  $h_{ab}=g_{ab}-n_a n_b$ is the induced metric of the surface layer,  ${\tensor{\overline{\nabla}}{^a}}$ is the tangential covariant derivative restricted to the hypersurface, $[\tensor{C}{_a_b}]=\tensor{C}{^+_a_b}-\tensor{C}{^-_a_b}$ and $\tensor{\mathcal{G}}{_c_d}$ is the singular part of the Einstein tensor considered in a distributional sense.

\subsection{Application to the LTB metric}

Applying to the LTB metric the Lanczos-Israel-formalism yields as the only nonzero component of the Einstein tensor: $\tensor{G}{_t_t}$, given by \eqref{ypunto}, while the extrinsic curvature at the hypersurface marked locally by $r=\pi/2$ is given by
\begin{equation}\label{extcurvgen}
\tensor{K}{^a_b}=\left[\begin{array}{c c c c}
0&0&0&0\\
0&0&0&0\\
0&0&-\frac{\sqrt{1-K}|R'|}{R R'}&0\\
0&0&0&-\frac{\sqrt{1-K}|R'|}{R R'}
\end{array} \right].
\end{equation}
The Darmois junction conditions demand the continuity of $\tensor{K}{^a_b}$ at the hypersurface. Hence, if $R'>0$ for all $r$, then $|R'|/R'=1$, so that  the junction conditions are equivalent to the continuity of $R$ and $K$. If there exists a zero of $R'$ in some fixed value $r=r_0$, then there exists a discontinuity of $\tensor{K}{^a_b}$ unless $K(r_0)=1$. At the turning value  $r_0=r^*=\pi/2$ there is clearly a discontinuity of $\tensor{K}{^a_b}$.

Bonnor proved that a PACM must be an elliptic model. First, he proved that if $R'$ changes sign (turning value) on a hypersurface $r=r^*$, with $r^*$ constant, and $1-K\neq 0$ on the hypersurface, then there is a surface layer. The proof is straightforward. Since $R$ has two zeros (two symmetry centers) in closed LTB models, the continuity of $R$ implies the existence of a turning value marked by a zero of $R'$ in some value $r=r^*$ within the radial coordinate range between the centers. Bonnor's condition \ref{condi2} implies that $r=r^*$ must lie within an elliptic region ($K>0$), since the regularity condition  $1-K=0$ at $r=r^*$ cannot be satisfied for a turning value in parabolic or hyperbolic regions ($K\leq 0$). Turning values in such regions necessarily exhibit a surface layer, which is not contemplated in the definition of a PACM.

The equation of state of the surface layer that follows from (\ref{distribfe}) is $\sigma+\Pi_1+\Pi_2=0$, where $\sigma$ is the surface density and $\Pi_i$ are the surface pressures that follow from the right hand side of (\ref{distribfe}) (the distributional energy--momentum tensor). Bonnor considered this equation of state unphysical, not only for having negative pressure, but also because of:
$$M_{TS}=\int (\tensor{T}{^1_1}+\tensor{T}{^2_2}+\tensor{T}{^3_3}-\tensor{T}{^4_4})\sqrt{h}\; d^3x=\int(\sigma+\Pi_1+\Pi_2)\sqrt{h}\; d^3x=0,$$
which means that the surface layer energy--momentum tensor produces zero active gravitational mass.  

To choose the appropriate free functions $M,\,K,\,t_{bb}$ for a closed model we must demand that their radial gradients vanish at turning values and at the symmetry centers. In the following sections we re-examine and extend Bonnor's results, looking at the spatially flat ($K=0$) and negatively curved ($K<0$) cases separately. 

\subsection{Surface tension}\label{sec:surften}

The presence of distributional sources in thin layers can be associated with surface tension through the relativistic generalisation of the Kelvin relation of Newtonian physics \cite{Schmidt} 
\begin{equation}\label{eqkelvin}
\Delta P = - 2{\cal K}A
\end{equation}
where the surface tension $A$ depends on the material, $\Delta P$ the difference of pressures in both sides of the surface layer and ${\cal K}$ is the mean curvature given by ${\cal K}=1/R_1+1/R_2$, with $R_1$, $R_2$ the principal curvature radii. As proven in \cite{Schmidt}, the relativistic  generalisation of \eqref{eqkelvin} is connected to a thin shell in the framework of the Israel--Lanczos formalism: 
\begin{equation}\label{eqkelg}
\Delta P=\frac{1}{2}\left(\tensor{K}{^+_\alpha_\beta}+\tensor{K}{^-_\alpha_\beta} \right)\tensor{\mathcal{T}}{^\alpha^\beta}.
\end{equation}
where $\tensor{\mathcal{T}}{^\alpha ^\beta}$ is the projected energy--momentum tensor in (\ref{distribfe})
\begin{equation}
\tensor{\mathcal{T}}{^\alpha ^\beta}=h^\alpha_a h^\beta_b \tensor{\mathcal{T}}{^a^b},\qquad   8\pi\tensor{\mathcal{T}}{^a^b}=-[\tensor{K}{^a ^b}]+\tensor{h}{^a ^b}[\tensor{K}{^c_c}]  \end{equation}
where $h^\alpha_a = \delta^\alpha_a\delta^\beta_b+n^\alpha_a n^\beta_b$ with $\alpha, \beta=t,\theta,\phi$ the hypersurface intrinsic coordinates.    

 \section{The spatially flat case $K=0$}\label{sec:seckzero}

A convenient choice for the free functions $M$ and $t_{bb}$ is
\begin{equation}\label{elecfunc}
M=M_0 \sin^3 \bar{r},\quad t_{bb}=-T_0 \sin^2 \bar{r}, \quad \Rightarrow \quad R=\left(\frac{9}{2}M_0\right)^{1/3} \sin \bar{r}\;\left[\bar{t}+T_0\sin^2 \bar{r}\right]^{2/3},
\end{equation}
where $M_0=\frac32 H_0^{-1}$, $T_0$ is an arbitrary constant, $\bar{r}=\pi H_0 r$ and $\bar{t}=H_0 t$ are the radial and time dimensionless coordinates respectively. However, to simply notation henceforth we will drop the bars on top of $t$ and $r$, understanding henceforth that (unless specifically stated) $t$ and $r$ without overbars denote these dimensionless rescaled coordinates.

The parameters in (\ref{elecfunc}) have been selected so that the kinematic evolution of the model at the symmetry centres $r=0,\,\pi$ coincides with that of the Einstein--de Sitter spatially flat FLRW model, whose Big Bang time is given by $t=0$. Hence, the constant $T_0$ can be identified with the Big Bang time of the LTB model at $\bar{r}=\pi/2$ (or equivalently $r=\frac12 H_0^{-1}$), that is: $t_{bb}(\pi/2)=-T_0<0$. For a more realistic cosmological scenario in the context of an inhomogeneous model with small deviation from an FLRW background, we shall assume that $|T_0|\ll t_0$, with present cosmic age given by $t_0\sim 13.7 \times 10^9$ years (a convenient bound value is $|T_0|\sim 10,000$ years).  
With this choice of free functions we have 
$$R'=\frac{(M_0/6)^{1/3}\cos r\, \left[ 7 \,T_0\sin^2 r+3t\right]}{\left[ t+T_0\sin^2 r\right]^{1/3}}$$
while the density and the components of the extrinsic curvature follows from (\ref{ypunto}) and (\ref{extcurvgen}) for $K=0$:
\begin{equation}\label{rhok0}
8\pi\rho=\frac{16}{3(t+T_0\sin^2 r)(7T_0\sin^2 r+3t)}.
\end{equation}
\begin{eqnarray*}
\tensor{K}{^\theta_\theta}=\tensor{K}{^\phi_\phi}&=&-\frac{2}{3}\frac{{\cal H}\left(r-\frac{\pi}{2}\right)}{M_0^\frac{1}{3}\sin r\; \left[t+T_0\sin^2 r\right]^{2/3}},\\
\end{eqnarray*}
where ${\cal H}(r)$ is the Heaviside function and we used the fact that $t+t_0\sin^2 r\geq 0$ in the full domain $0\leq r\leq \pi$.
 
Since these expressions allow us to compute $\tensor{K}{_a_b}^++\tensor{K}{_a_b}^-=0$, while $\tensor{G}{_a_b}$ is continuous on $\mathcal{S}$, then \eqref{eqsenov1} is satisfied identically everywhere. On the other hand, the right hand side of \eqref{eqsenov2} is zero, but computing its covariant derivative and evaluating on $\mathcal{S}$ yields the following result:  the singular part of the Einstein tensor, $\tensor{\mathcal{G}}{_a_b}$, is constant on $\mathcal{S}$. Notice that from \eqref{eqkelg} there is no surface pressure due to surface tension.

At $\mathcal{S}$ the only nonzero components of the distributional energy-momentum tensor are:
$$8\pi \sigma=\frac{1}{(36 M_0)^{1/3}(t+T_0)^{2/3}},\;\;\;\; 8\pi\Pi_1=8\pi\Pi_2=-4\pi\sigma,$$
where $\sigma$ is the distributional density, while $\Pi_1$ and $\Pi_2$ are the distributional pressures, with the equation of state given (as found by Bonnor) by $\sigma+\Pi_1+\Pi_2=0$. As the units of the distributional and continuous (non--distributional) density are not the same we obtain the quasi-local mass from each density to obtain a quantity that can be compared. The energy momentum tensor is divided into a continuous (non-distributional) part and a distributional part as: $\overline{T}_{ab}=T_{ab}+\mathcal{T}_{ab}\delta(\mathcal{S})$, where $\delta$ is the Dirac delta function, and in our case $T_{ab}=\rho u_a u_b$. From the expression of the full energy-momentum tensor it is clear that it makes sense to compare the quantities $ T_{ab}u^a u^b$ and $\mathcal{T}_{ab}u^a u^b\delta(\mathcal{S})$, which have the same energy density units, by means of integration over a domain that contains the hypersurface.

We integrate $\rho=T_{ab}u^a u^b$ in a domain $0<r_1<\pi/2<r_2<\pi$,  
\begin{equation}\label{intrho}
\mathcal{M}_{\rho}=4\pi\int_{r_1}^{r_2}\rho R^2 R' dr= M_0 [ \sin^3 r_2-\sin^3 r_1],
\end{equation}
from this expression we obtain an upper and lower bound, $0\leq \mathcal{M}_{\rho}\leq M_0$.

For the distributional matter at the thin shell we obtain the contribution of $\sigma$ to the active gravitational mass as the integral of $\mathcal{T}_{ab}u^a u^b\delta(\mathcal{S})$,
\begin{align*}
\mathcal{M}_{\sigma}&= \int_{r_1}^{r_2}\sigma R^2\vert_{\mathcal{S}}\delta\left(r-\frac{\pi}{2}\right) \int d\Omega\; dr = \frac{1}{2} \int_{r_1}^{r_2} \frac{\left(\frac{9}{2}M_0\right)^{\frac{2}{3}} \;  (t+T_0)^\frac{4}{3}}{6^\frac{2}{3}M_0^\frac{1}{3}(t+T_0)^\frac{2}{3}}\delta\left(r-\frac{\pi}{2}\right) \; dr\\
&= \frac{1}{2}\left(\frac{9 M_0}{16}\right)^{1/3}(t+T_0)^{2/3}.
\end{align*}
Considering the arbitrary $0\leq r_1,\,r_2\leq \pi$ which give the upper bound for $\mathcal{M}_{\rho}$, and from the ratio of the latter and $\mathcal{M}_{\sigma}$ we obtain a comparison of the continuous mass and the contribution of the distributional density to the quasilocal mass 
\begin{equation}\label{cocmass}
\xi(t)=\frac{\mathcal{M}_{\rho}}{\mathcal{M}_{\sigma}} = 2\left(\frac{4}{3}\right)^\frac{2}{3} \frac{M_0^\frac{2}{3}}{(t+T_0)^\frac{2}{3}}.
\end{equation}
To obtain a numerical result we evaluate this ratio at present day cosmic time $t_0\approx 13.7 \times 10^9$ years and use $M_0= 3/2 H_0^{-1}$, where $H_0$ is the Hubble constant ($\sim 70\hbox{km}/(\hbox{sec}\,\hbox{Mpc})$). We obtain for these values
\begin{equation}
\xi(t_0)\approx \frac{2^\frac{2}{3} 2}{H_0^\frac{2}{3}( t_0 + T_0)^\frac{2}{3}} \approx  \frac{2^\frac{2}{3} 2}{\left(70 \frac{\textrm{km/s}}{\textrm{Mpc}}\right)(13.7\times 10^9+10^5)\textrm{years}} \approx 3.2176,
\end{equation}
while for ten times the current age of the universe we have
\begin{equation}
\xi(10\,t_0)\approx  \frac{2^\frac{5}{3}}{\left(70 \frac{\textrm{km/s}}{\textrm{Mpc}}\right)(13.7\times 10^{10}+10^5)\textrm{years}} \approx 0.69.
\end{equation}
Thus, for the asymptotic evolution range of large cosmic times the contribution to the quasi--local mass from the distributional surface density dominates the contribution from the continuous dust source. This behaviour is clearly unphysical, since the distributional source does not generate effective gravitational mass (from the quasi--local mass definition), yet it ends up overwhelmingly dominating over the quasilocal mass obtained from the continuous (and physical) dust density. In section \ref{sec:geod} we further examine the physical implications of this model.

\section{The case $K<0$}

We select the same free functions as in \eqref{elecfunc}, together with $K(r)=-K_0\sin^2 r$. The only non-vanishing components of the extrinsic curvature are
\begin{eqnarray*}
\tensor{K}{_\theta_\theta}=-\frac{\sqrt{1-K_0\sin^2 r} R |R'|}{R'}, \;\;\;\;\; \tensor{K}{_\phi_\phi}=\tensor{K}{_\theta_\theta}\sin^2\theta,
\end{eqnarray*}
where
$$R(t,r)=\frac{M_0\sin r (\cosh\eta -1)}{K_0},\;\;\;\;\; \eta-\sinh\eta=\frac{K_0^\frac{3}{2}(t+t_0\sin^2 r)}{M_0}.$$
Once again, $[\tensor{\underline{G}}{^a_b}]=0$, and $\tensor{K}{_a_b}^++\tensor{K}{_a_b}^-=0$, so \eqref{eqsenov1} is satisfied. Taking the covariant derivative of $\tensor{\mathcal{G}}{_a_b}$ on $\mathcal{S}$ leads to a zero vector and thus \eqref{eqsenov2} is identically satisfied once again. At $\mathcal{S}$ the distributional density and pressures are
$$8\pi \sigma=4\frac{\sqrt{1-K_0}}{R\left(t,r\right)|_{r=\frac{1}{2H_0}}},\;\;\;\; 8\pi\Pi_1= 8\pi\Pi_2 = -4\pi \sigma,$$
while the non-distributional density takes the form
\begin{equation}
8\pi \rho=\frac{3M_0 \sin^2 r \cos r}{4\pi R^2 R'}.
\end{equation}
To obtain a comparison one would proceed as in the case $K=0$ but taking into account the proper mass instead of the quasilocal mass, as in this case both masses are not equal. These comparison yields a similar result as in the case studied previously, which we considered to be unphysical. 

\section{Case $\Lambda >0$}\label{sec:lamg0}

If $\Lambda >0$, Einstein's field equations yield the same form for the density $\rho$ given in (\ref{ypunto}), but the Friedman--like evolution equation is now:
\begin{equation}\label{ydotlam}
\dot{R}^2=\frac{Q(R)}{R},\qquad Q(R)=2M - K R+\lambda R^3
\end{equation}
where $\lambda=\frac{1}{3}\Lambda$. The kinematic evolution is governed by the zeroes of the cubic polynomial $Q(R)$ for different values of $K$. Ever expanding regions or models are characterised by configurations with those choices $K$ and $M$ for which $Q$ has no zeros for a specific range of $r$. In particular, fully regular closed ever expanding models without thin layer distributional sources require configurations with $K>0$ for which $Q(R)$ has no zeroes for all the range of $r$.   

\begin{figure}[H]
\centering
  \includegraphics[width=2.6in]{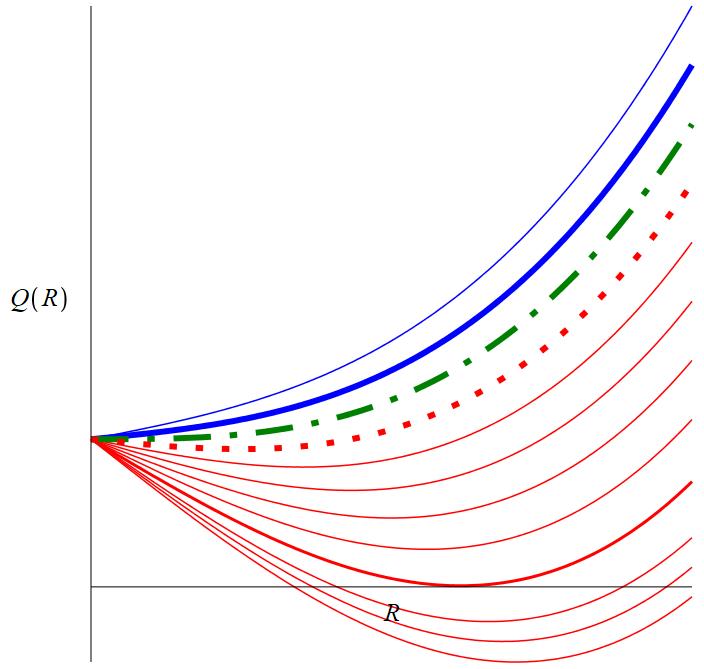}
  \centering\caption{Plot of Q(R) in \eqref{ydotlam}, see text for explanation.}
  \label{fig:polyfig}
\end{figure}
To look at the sign of $Q(R)$ we plot this cubic polynomial for fixed positive values of $M$ and $\lambda$ and letting vary $K$ for $R>0$. As shown in Fig. \ref{fig:polyfig}, all curves above the lowest red thick curve (colors appear in the online version), which are configurations of a generic LTB dust solution, represent ever expanding universes. The dot-dash green curve represents spatially flat models, below this curve are models with $K>0$, and above the dash-dot green curve there are negative spatial curvature models. In this case we can choose $K>0$ so that the condition $K(r^*)=1$ for $R'(t,r^*)=0$ holds and thus, we have ever expanding models for which the regularity conditions for a PACM hold: the metric coefficient $\sqrt{g_{rr}}=\pm R'/\sqrt{1-K}$ is well defined at $r^*$ and $K^a_b$ is continuous, which eliminates the surface distributional source at $r=r^*$. This is an important result, since it proves that LTB models that approximate the $\Lambda$-CDM model can have rest frames with a closed topology.

\section{Radial null geodesics at the interface}\label{sec:geod}

While the thin shell distributional source at the hypersurface $r=r^*$ in ever expanding closed LTB models does not generate effective mass, it is interesting to find out if the existence of such source could be detected observationally. To explore this question we need to find null geodesics that cross this hypersurface and compute the redshift from light emitted along these curves by distant observers in these models. 

Photon trajectories (null geodesics) follow from the solutions of the geodesic equation,
\begin{equation}\label{ecgeodesica}
\frac{d^2 \tensor{x}{^a}}{d\lambda^2}+\tensor{\Gamma}{^a_b_c}\frac{d \tensor{x}{^b}}{d\lambda}\frac{d \tensor{x}{^c}}{d\lambda}=0,
\end{equation}
with the constraint $k^a k_a=0$, for $k^a=dx^a/d\lambda$ is the tangent vector of these curves and $\lambda$ is an affine parameter. We will consider only radial null geodesics $k^a=[k^t(\lambda),k^r(\lambda),0,0]$, where $k^t$ and $k^r$ are obtained from (\ref{ecgeodesica})

\begin{align}
 \frac{d^2 t}{d\lambda^2}+\frac{\dot{R}'}{R'}\left( \frac{dt}{d\lambda}\right)^2&=0,\label{nulradt}\\
 \frac{d^2 r}{d\lambda^2}+\left(\frac{R''}{R'}-\frac{K'}{2(1-K)}\right)\left( \frac{dr}{d\lambda}\right)^2\pm \frac{2\dot{R}'}{\sqrt{1-K}}\frac{|R'|}{R'}\left(\frac{dr}{d\lambda}\right)^2&=0\label{nulradr},
\end{align}
subjected to the constraint $k_ak^a=0$
\begin{equation}
-\left(\frac{dt}{d\lambda}\right)^2+\frac{R'^2}{1-K}\left(\frac{dr}{d\lambda}\right)^2=0, \;\; \Rightarrow \;\; \frac{dt}{d\lambda}=\pm \frac{R'}{\sqrt{1-K}} \frac{dr}{d\lambda}.\label{eqconstplus}
\end{equation}
The metric functions $R,\,K$ and their derivatives in the coefficients follow from the closed ever expanding models we have examined in previous sections (with $\Lambda=0$). 

It is well-known that a non-degenerate $C^{r+1}$ metric determines the $C^r$ Levi-Civita connection. For $K=0$ the metric is $C^\infty$, for $K\neq 0$ in general it can only state that the metric is $C^0$. For convinience we will analyze the case $K=0$ in which the connection is $C^r$ almost everywhere, i.e. it is $C^r$ except on a set of measure zero, namely the symmetry centers and at the turning value of $R'$. Therefore there exists a convex normal neighborhood at each $p\in M$, i.e. an open set $U$ with $p\in U$ such that for all $q,r\in U$ there exists a unique geodesic $\gamma$ which stars at $q$ and ends at $r$ and is totally contained in $U$, see \cite{Hawking}. The connection is not $C^r$ at the symmetry centers and at the hypersurface $r=r_*$, nevertheless the radial geodesic equation is $C^r$ in all the space-time except at the hypersurface $r=r_*$. By the standard existence and uniqueness theorem for ODE's there exists a unique geodesic from a symmetry point to any point arbitrarily near the hypersurface, in comoving coordinates this guarantees the existence of a null geodesic that starts at $r=0$ and ends at $r=r_*-\epsilon_1$ for any $\epsilon_1>0$ and a null geodesic with endpoints at $r=r_*+\epsilon_2$ and $r=r_c$ for all $\epsilon_2>0$.

In order to check if the geodesic equation is well defined at $r=\pi/2$, we consider the choice of functions of section \ref{sec:seckzero}, leading to: 
\begin{align}
\frac{d^2 t}{d\lambda^2}+\frac{2}{3}\Phi(t,r)\left(\frac{d t}{d\lambda}\right)^2=&0,\label{eqtgeoK0}\\
\frac{d^2 r}{d\lambda^2}+\frac{\Psi(t,r)-\Omega(t,r)}{21\cos r\left(t_0\sin^2 r+\frac{3}{7}t\right)(t+t_0\sin^2 r )}\Psi(t,r)\left(\frac{dr}{d\lambda}\right)^2=&0\label{eqrgeoK0}.
\end{align}
where
\begin{align}
\Phi(t,r)=&\frac{3t+t_0\sin^2 r}{(t+t_0\sin^2 r )\left(3t+7t_0\sin^2 r \right)},\\
\Psi(t,r)=&\pm \frac{2(36M_0)^\frac{1}{3}}{3}\cos r (t_0\sin^2 r +3t)\left\vert \frac{\cos r(3t+7t_0\sin^2 r )}{(t+t_0\sin^2 r )^\frac{1}{3}}\right\vert,\label{eqPsi}\\
\Omega(t,r)=&21\left(t_0^2\sin^4 r +\left(\frac{10}{7}tt_0-\frac{4}{3}\cos^2 r \right)\sin^2 r+\frac{3}{7}t(t-4t_0\cos^2 r )\right)\sin r,
\end{align}
and the plus minus sign in the square root from equation \eqref{eqconstplus} will distinguish between ``ingoing'' past directed curves and ``outgoing'' future directed curves. 

\begin{figure}[H]
\centering
  \includegraphics[width=4in]{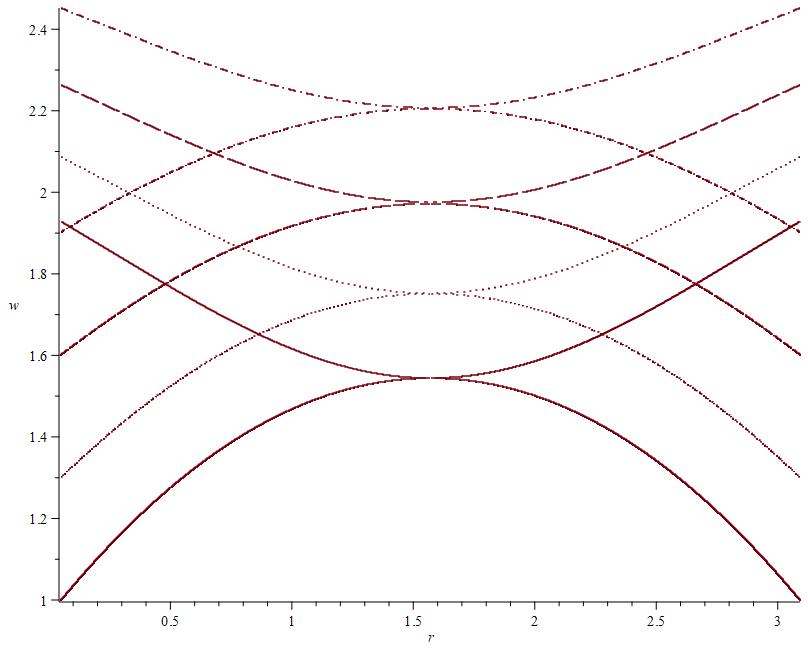}
  \centering\caption{Plot to numerical solution of equations \eqref{eqtgeoK0} and \eqref{eqrgeoK0} for four different geodesics with generic initial conditions. As it can be noticed, the curves are smooth when considered as $w(r)$ as opposed to the discontinuity that is examined when $w$ and $r$ are considered as functions of the parameter $\lambda$.}
  \label{fig:plotode}
\end{figure}

Since \eqref{eqrgeoK0} is not well-defined near $r(\lambda)=\pi/2$, we introduce the change of variable: $t(\lambda)=10^{w(\lambda)}$ and solve numerically the geodesic equations above for generic values of $M_0$ and $t_0$. In what follows we consider  $M_0=10$ and $t_0=0.5$. The absolute value needs to be evaluated in a piecewise manner $|x|=x$ for $x>0$ and $|x|=-x$ for $x<0$ for any $x$. For generic initial conditions and working with both signs, each considered also in the geodesic equations (see \eqref{eqPsi}) we solve numerically \eqref{eqtgeoK0} and \eqref{eqrgeoK0} for several initial conditions, leading to the curves plotted in figure \ref{fig:plotode}.

\begin{figure}
\hfill
\subfigure[Plot for $r(\lambda)\in [0,\pi/2)$]{\includegraphics[width=6cm]{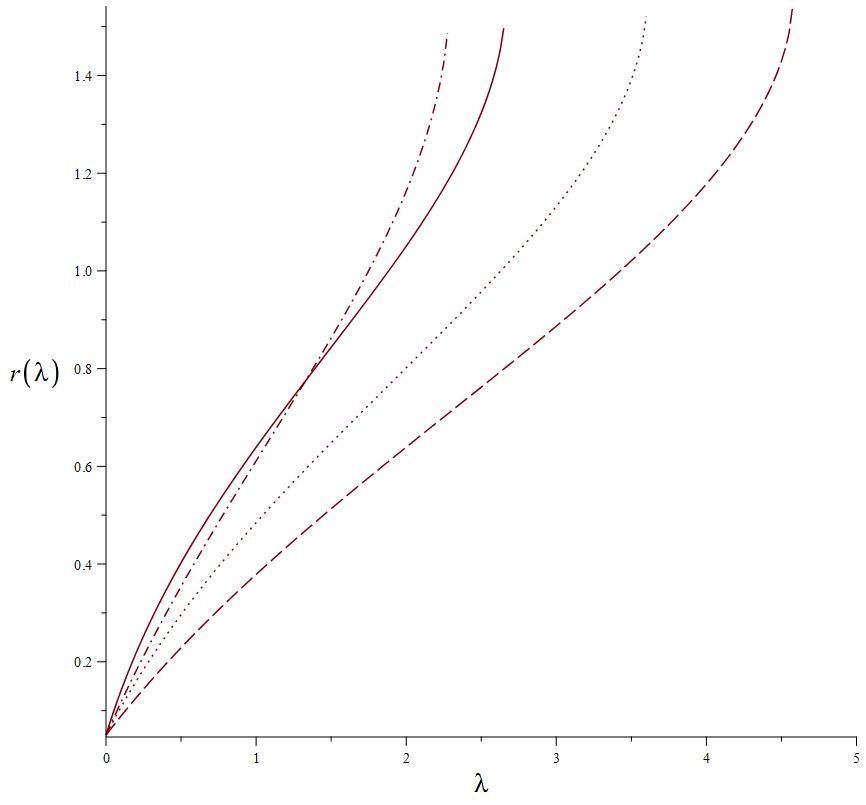}}
\hfill
\subfigure[Plot for $r(\lambda)\in {\left(\pi/2, \pi \right]} $]{\includegraphics[width=6cm]{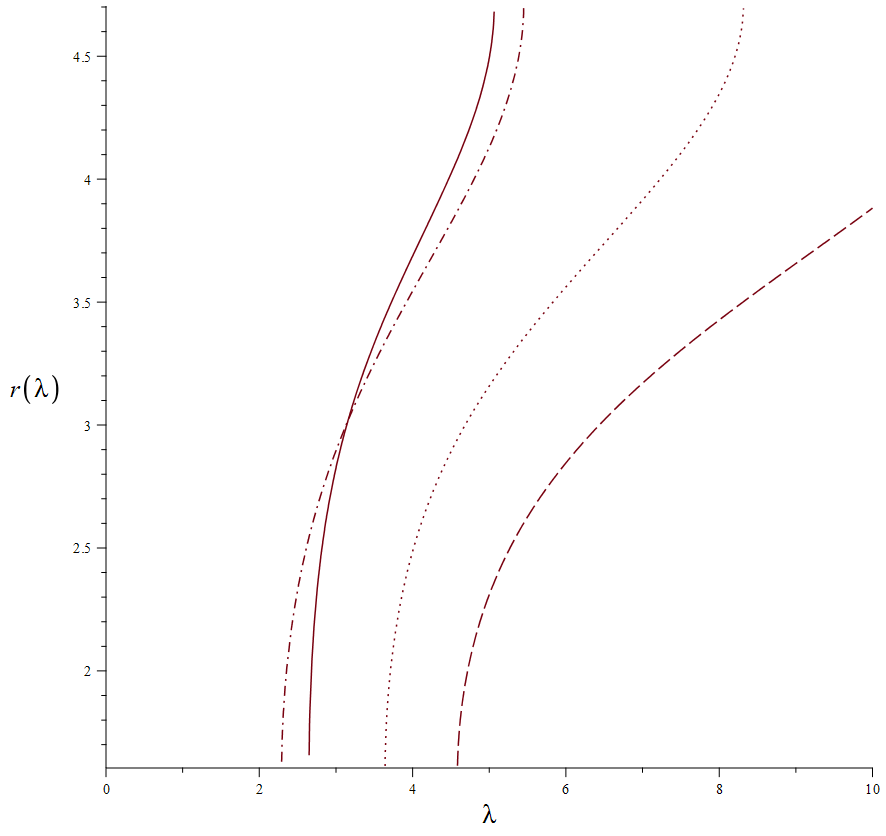}}
\hfill
\caption{Plot to numerical solutions for $r(\lambda)$. The plot on the left is the plot for the values of $r\in [0,\pi/2)$, while the second graph represents the values $r\in(\pi/2,\pi]$.  From the plots it can be seen that the values of the derivative $dr/d\lambda$ diverge, which has implications on $dt/d\lambda$, see fig. \ref{fig:odesolplott}.}\label{fig:odesolplot}
\end{figure}

 The numerical solution for $r\in\left[0,\pi/2\right)$ shows that near $\pi/2$ the derivative $dr/d\lambda$ does not tend to zero. The graphs for $r(\lambda)$ and $t(\lambda)$ for some of the geodesics obtained are shown in figure \ref{fig:odesolplot}. It can be seen from the solutions that \eqref{eqconstplus} restricts the solutions for $t(\lambda)$ and $r(\lambda)$ to be such that the product $R' dr/d\lambda$ be finite. In this cases the product is not zero which implies that $dr/d\lambda$ must diverge. Also, equation \eqref{eqconstplus} reveals that solutions that are not $C^1$ can be obtained, as arbitrary initial conditions can be chosen over $r$ as a function of $\lambda$ to obtain a $C^0$ curve, defining $r(\pi/2)=\lim_{r\to {\pi/2}^+}r(\lambda)=\lim_{r\to {\pi/2}^-}r(\lambda)$ that satisfies \eqref{eqtgeoK0} and \eqref{eqrgeoK0} for $r\in [0,\pi/2) \cup (\pi/2,\pi]$. Some of these solutions are shown in figure \ref{fig:plotode}. Therefore, there exists a jump in the first derivative of the curve which could be used to probe the existence of thin shells.

\begin{figure}
\centering
  \includegraphics[width=3.5in]{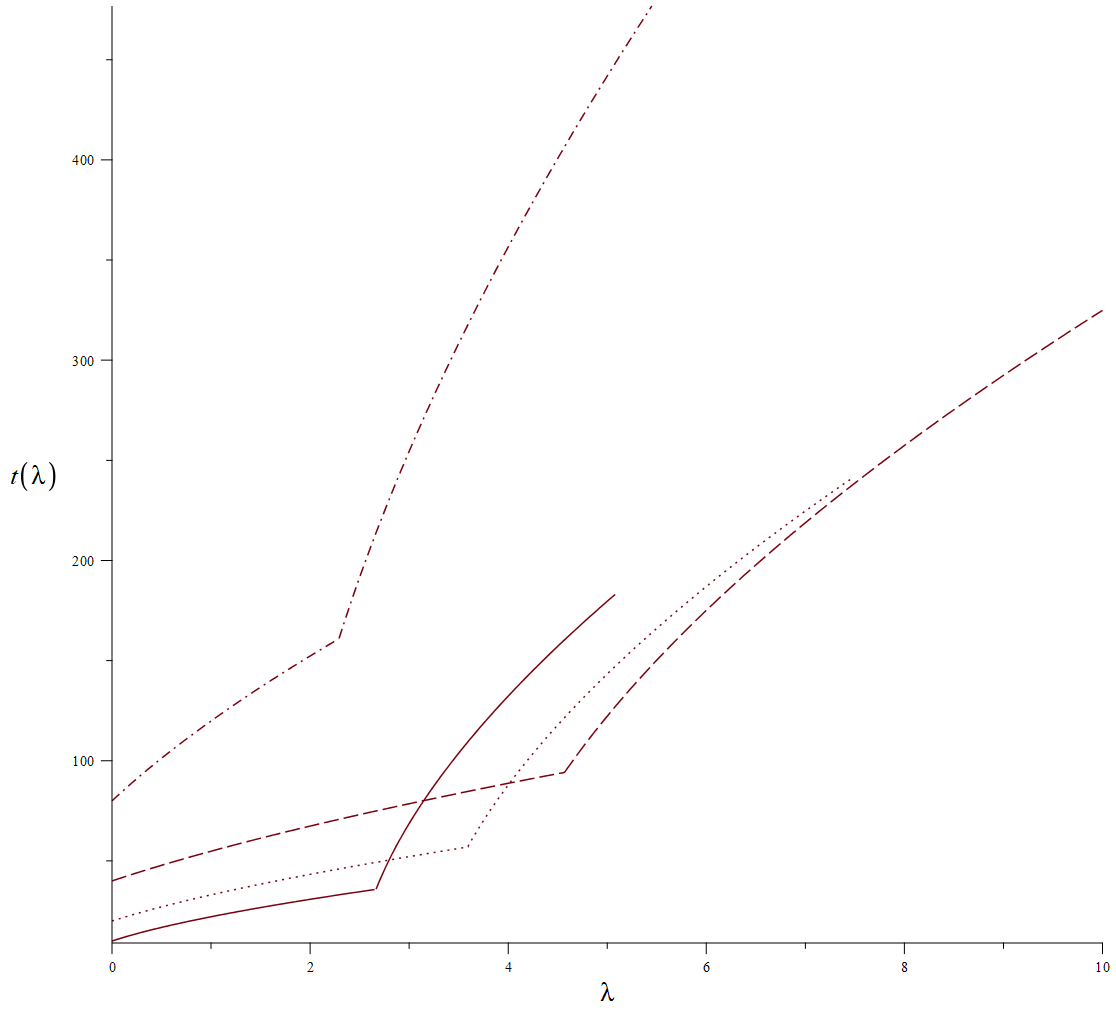}
  \centering\caption{Plot to numerical solutions for $t$, from the plots it can be seen that the values of the derivative of the $t$ curve are always finite due to the divergence of $dr/d\lambda$, see fig. \ref{fig:odesolplot}, as required by equation \eqref{eqconstplus}.}
  \label{fig:odesolplott}
\end{figure}
Although there is a discontinuity in the first derivative of the coordinates of the geodesics, each value of $r\in [0,\pi/2)\cup(\pi/2,\pi]$ is reached in a finite value of the affine parameter.

\section{Redshift}\label{sec:redshift}

\begin{figure}
\hfill
\subfigure[Plot to numerical solution for $z(\lambda)$]{\includegraphics[width=6cm]{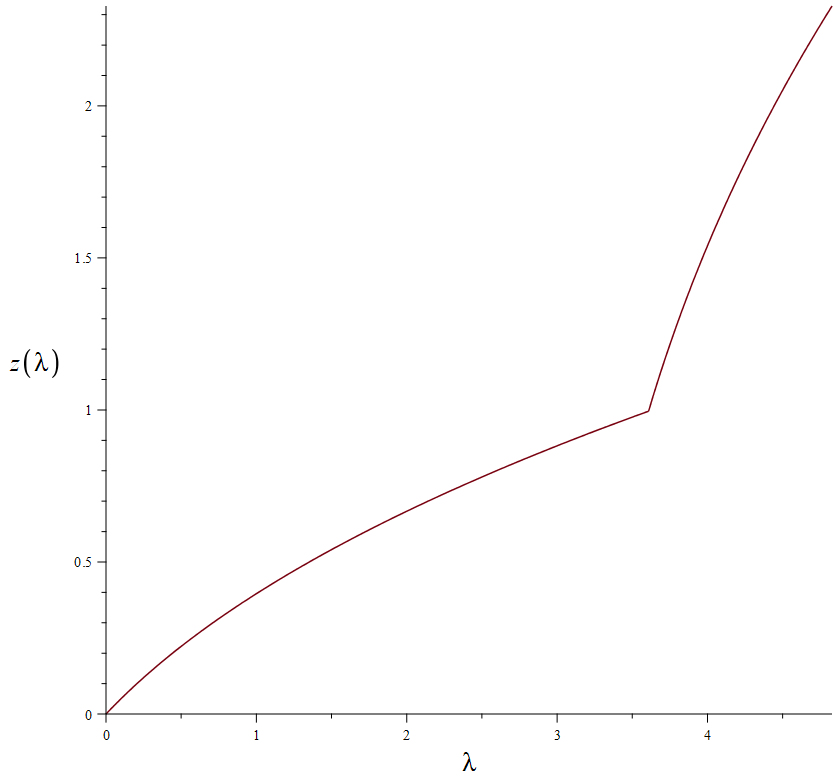}}
\hfill
\subfigure[Plot to numerical solution for $\frac{1}{1+z}\frac{dz}{d\lambda}$]{\includegraphics[width=6cm]{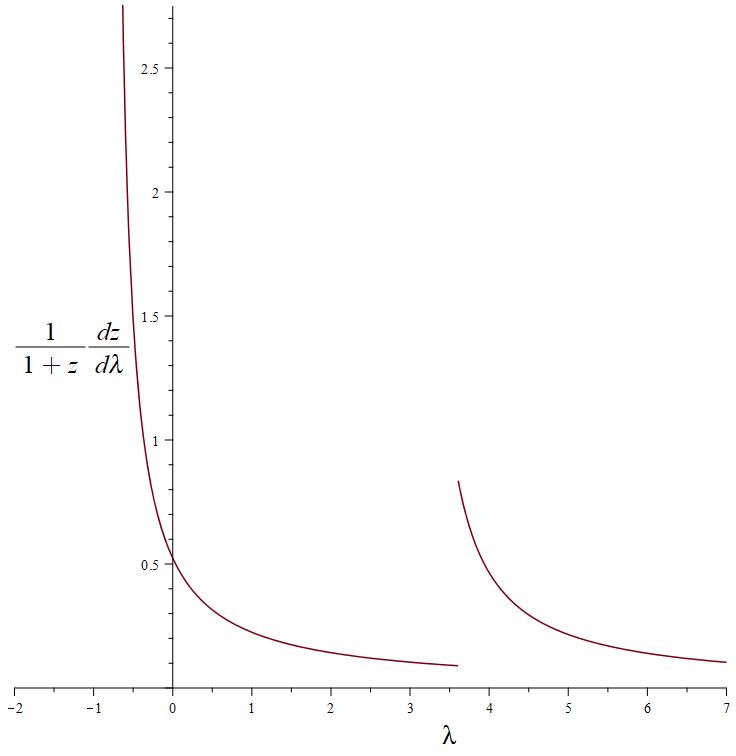}}
\hfill
\caption{Plot to numerical solutions for first geodesic, where the disconinuity of the redshift can be apreciated. See text for details.}\label{fig:redplot1}
\end{figure}

\begin{figure}
\hfill
\subfigure[Plot to numerical solution for $z(\lambda)$]{\includegraphics[width=6cm]{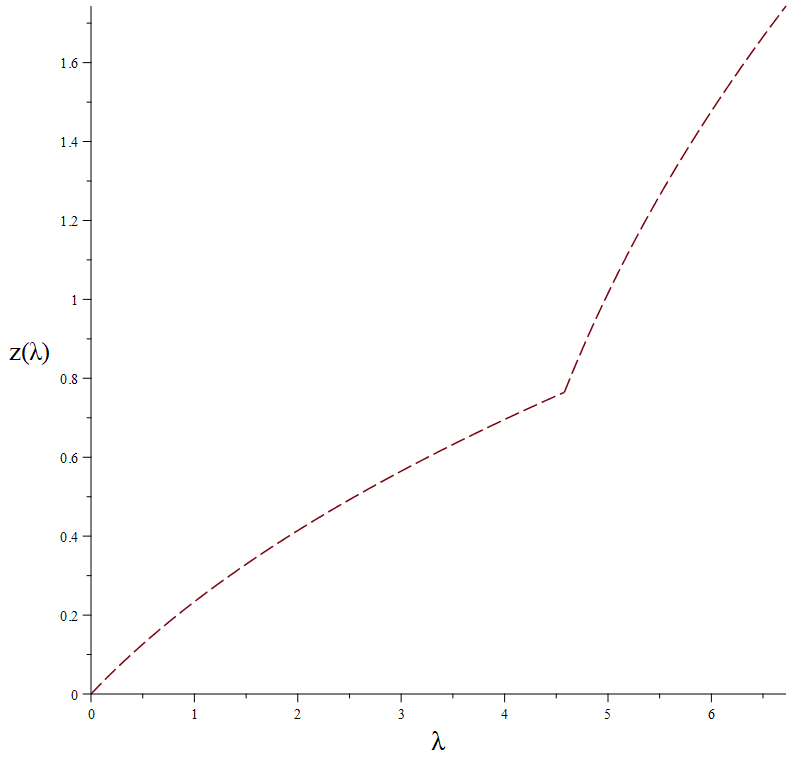}}
\hfill
\subfigure[Plot to numerical solution for $\frac{1}{1+z}\frac{dz}{d\lambda}$]{\includegraphics[width=6cm]{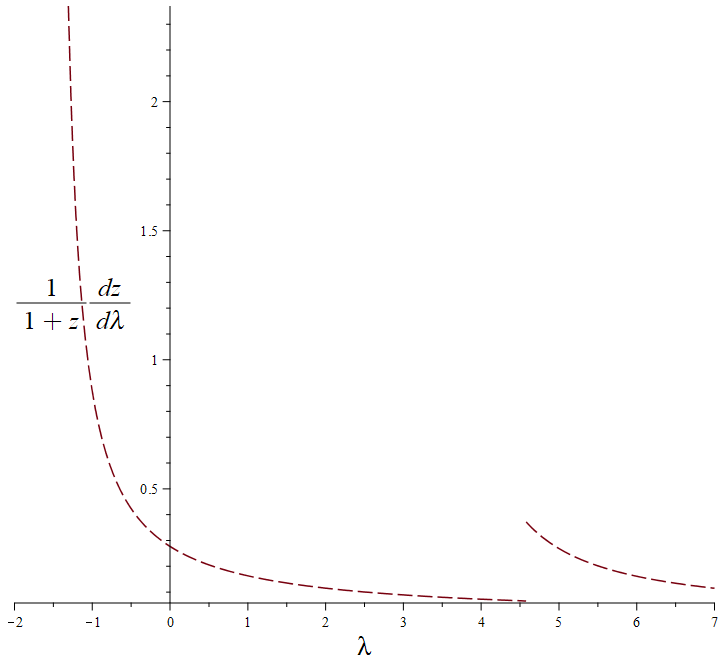}}
\hfill
\caption{Plot to numerical solutions for second geodesic, where the disconinuity of the redshift can be apreciated. See text for details.}\label{fig:redplot2}
\end{figure}

The redshift for a $K=0$ model is calculated through the following integral \cite{Bolejko}
\begin{equation}\label{redshiftlambda}
\ln(1+z(r(\lambda)))=\int_0^{\lambda} \dot{R}'(t(\lambda),r(\lambda))\; \frac{dr}{d\lambda} d\lambda.
\end{equation}
Note that as $dr/d\lambda$ is discontinuous at $r=\pi/2$, the integrand is not continuous but the integral is. Figures \ref{fig:redplot1} and \ref{fig:redplot2} represent the redshift and the plot for $1/(1+z) dz/d\lambda$ for two different geodesics.\\

As there is a discontinuity in the derivative of the redshift it is possible probe the existence of a thin shell by measuring the redshift of radial photons that cross the surface $\bar r=\pi/2$ (which corresponds to a physical comoving distance $r=1/(2H_0)$). Nevertheless notice that the magnitude of the discontinuity depends on the parametrization chosen.
\section{A model with $K>0$}\label{sec:Kpos}
We now analyze the case $K>0$ and show that in this case observers at the turning value would not detect any thin layers. Analyzing a model with positive $K$ is easier with a change of variables in the metric, where we obtain a FLRW-like metric with line element
\begin{equation}\label{flrw}
ds^2=-dt^2+a^2 \left[\frac{\Gamma^2 R_i'^2}{1-k_{qi}R_i^2}dr^2+R_i^2 d\Omega^2 \right],
\end{equation}
where $a(t,r)\equiv R/R_i$, $R_i\equiv R(t_i,r)$ and $t=t_i$ determines a fiducial initial hypersurface. Henceforth, all quantities evaluated at $t=t_i$ will be denoted by a the subindex $i$. The dimensionless metric function $\Gamma$ is
$$\Gamma \equiv \frac{R'/R}{R_i'/R_i}=1+\frac{a'/a}{R_i'/R_i},$$
where $\Gamma_i=1$, while $k_{qi}=K/R_i^2$. Note that the regularity condition on this metric is $R'=0$ which implies $k_{qi}R_i^2=1$.

We now consider the functions $a,\Gamma$ and $R_i$ taking into account a closed model. We have, c.f. \cite{Sussman}, that along turning values regularity conditions on the density, $\rho$, density at the fiducial time, $\rho_i$, Ricci scalar of the hypersurfaces at the fiducial time $ {}^{(3)}\mathcal{R}_i$ and the metric imply that $R_i'$, $M'$ and $(k_i R_i)'$ must have common zeros along the turning values of the same order in $r-\pi/2$. The function $\Gamma$ must not have a zero due to the fact that $\Gamma=0$ and $\rho_i>0$ imply a shell-crossing singularity. So, taking note of these considerations we have
$$R'=a'R_i+\frac{RR_i'}{R_i},\;\;\;\; K'=k_{qi}'R_i^2+2k_{qi}R_iR_i',$$
which yield $a'=k_i'=0$ when evaluating both equations along the hypersurface $\mathcal{S}$. 

The null geodesic constraint is 
\begin{equation}\label{nullgeoflrw}
\frac{dt}{d\lambda}=\pm a \frac{\Gamma R_i'}{\sqrt{1-k_{qi}R_i^2}}\frac{dr}{d\lambda},
\end{equation}
while the radial null geodesic equations are
$$\frac{d^2t}{d\lambda^2}+ \frac{a^2\dot{\Gamma}+a\dot{a}\Gamma}{2\Gamma^2}\left(\frac{dt}{d\lambda}\right)^2 =0,\;\;\;\; \frac{d^2 r}{d\lambda^2}+A(B+D)  \left(\frac{dr}{d\lambda}\right)^2 =0,$$
where
\begin{align*}
A&=\frac{1}{2a R_i'^2\Gamma(-1+k_{qi}R_i^2)},\\
B&=
2aR_i'\left[C\Gamma+\left(\pm\frac{\Gamma R_i'^2\dot{\Gamma}}{\sqrt{1-k_{qi}R_i^2}}+\frac{1}{2}R_i'\Gamma'\right)(-1+k_{qi}R_i^2)\right],\\
C&=-\frac{1}{2}k_{qi}'R_i'R_i^2-R_i'^2k_{qi}R_i+R_i''k_{qi}R_i^2-R_i',\\
D&=\left(2a'R_i'^2\Gamma\pm \frac{4\dot{a}R_i'^3\Gamma^2}{\sqrt{1-k_{qi}R_i^2}}\right)(-1+k_{qi}R_i^2).
\end{align*}
Next, we verify if the geodesic equation in these variables is well defined through the following limit
$$\lim_{r\to\frac{\pi}{2}}\Gamma=\lim_{r\to\frac{\pi}{2}}\frac{R'}{R_i'}\frac{R_i}{R}=\left(\lim_{r\to\frac{\pi}{2}}\frac{R'}{R_i'}\right)\frac{R_i}{R}\bigg\vert_{r=\frac{\pi}{2}}.$$
Since $R_i'=dR_i/dr$ is the derivative of a radial profile at a given time, and the partial derivative $R'=\partial R/\partial r$ is taken as a limit at a constant time, the limit in parenthesis above must be a finite function of time. We now check the product $AB$
\begin{equation}\label{AB}
AB=-\frac{C}{R_i'(1-k_{qi})}\pm \frac{R_i'\dot{\Gamma}}{\sqrt{1-k_{qi}R_i^2}}+\frac{1}{2}\frac{\Gamma'}{\Gamma}
\end{equation} 
The limit of the second term is  
$$\lim_{r\to\frac{\pi}{2}}\frac{R_i'\dot{\Gamma}}{\sqrt{1-k_{qi}R_i^2}}= \left(\lim_{r\to\frac{\pi}{2}}\dot{\Gamma}\right)\left(\lim_{r\to\frac{\pi}{2}}\frac{R_i'}{\sqrt{1-k_{qi}R_i^2}}\right).$$
The second limit of the right hand side is finite by regularity conditions, while the first 
$$\dot{\Gamma}= \left(\frac{\dot{R}'R+R'\dot{R}}{R^2}\right)\frac{R_i}{R_i'}= R_i\left(\frac{\dot{R}'R+R'\dot{R}}{R_i'R^2}\right)$$
has a finite limit at $r=\frac{\pi}{2}$ as long as the limit $\dot{R'}/R_i'$ exists. We now analyze the first term of the product $AB$,
\begin{align*} 
\frac{C}{R_i'(1-k_{qi}R_i^2)} = \frac{k_{qi}'R_i^2+2R_i'k_{qi}R_i}{2-2k_{qi}R_i^2} -\frac{R_i''}{R_i'}.
\end{align*}
The limit limit as $r\to\pi/2$ in the first term in the right hand side of the last equality does not exist, since $1-k_{qi}R_i^2$ has a zero of the same order as $R_i'^2$ and $k_{qi}'^2$. However, the non--existence of this is consistent with the geodesic equation at $\mathcal{S}$ not being defined in the comoving coordinates.

We now check the product $AD$, 
$$AD= \frac{a'}{a}\pm \frac{2\dot{a}R_i'\Gamma}{a\sqrt{1-k_{qi}R_i^2}}.$$
The limit of the first term in the right hand side is zero from the definition of $a$ and as $R'$ and $R_i'$ are continuous and zero at $\mathcal{S}$. The second term is constant by previous calculations.

We now compare the results for a model with $K>0$ with $K=K_0\sin(r)^2$. 
Regularity conditions for a closed model require that $K=1$ at $\mathcal{S}$ (where $R'=0$) and that the following limit be finite and nonzero:
\begin{equation}\label{limitreg}
\lim_{\mathbf{x}\to\mathcal{S}}\frac{R'^2}{1-K}, 
\end{equation}
where $\mathbf{x}$ denotes a generic point in the manifold. It is straightforward to prove that the metric component $g_{rr}$ is continuous but does not have a continuous partial derivative $g_{rr}'$ which immediately implies that the connection will not be $C^1$ in a set of measure zero, $\mathcal{S}$. On the contrary, in the case $K=0$ the connection is not $C^1$ due to the fact that the metric is degenerate at $\mathcal{S}$, as opposed to the model with $K>0$ which is not degenerate by regularity conditions.

Nevertheless, if a solution to the geodesic equation where to exist the derivative of the radial and temporal coordinates should be continuous as they must satisfy the null geodesic constraint \eqref{nullgeoflrw}, which relates both derivatives by the following relation 
\begin{equation}\label{constnullposK}
\frac{dt}{d\lambda}= \frac{\pm R'}{\sqrt{1-K}}\frac{dr}{d\lambda}.
\end{equation}
Both derivatives are related by a function which is continuous due to the regularity conditions, where it is used that the square root is a continuous function so the passage to the limit under the square root can be taken, and by hypothesis was assumed to be invertible, which completely determines both coordinates, unlike the case $K=0$ which gives an infinite number of choices of the derivative of the radial coordinate. Therefore LTB models with $K>0$ present no issues in geodesics and as there are no surface layers, and by the analysis of equation \eqref{formacero} there is no effect on the redshift nor on the derivatives of the coordinates.

\section{Conclusions and discussion}\label{sec:concl}

We have examined the dynamics and geometric properties of ever expanding ``closed'' LTB dust models, where by ``closed'' we mean models whose rest frames (hypersurfaces orthogonal to the 4--velocity marked by constant time) are diffeomorphic to the standard 3--sphere $\bf{S}^3$. We considered both cases, with $\Lambda=0$ and $\Lambda>0$. Since observations do not rule out a small positive curvature, the case $\Lambda>0$ can be thought of as a toy model inhomogeneous generalisation of the $\Lambda$CDM model. 

Ever expanding closed LTB models with $\Lambda=0$ where examined long time ago by Bonnor \cite{Bonnor}, who showed that fulfilment of regularity conditions require these models to admit a thin surface layer at the equator of the 3--sphere (``turning value'' of the area radius), which must be examined by means of the Israel--Lanczos thin shell formalism. Bonnor found the equation of state state satisfied by this distributional source, which he regarded as unphysical because it does not contribute to the effective quasi--local mass and because of the negative surface pressure (this was before negative pressures were acceptable in connection with dark energy).       

In the present article we extended Bonnor's work by looking at the time evolution of the distributional source, in comparison with the time evolution of the continuous dust source. We also show that assuming $\Lambda>0$ allows for perfectly regular closed LTB models, an option not contemplated by Bonnor. By looking first at the spatially flat case $K=0=\Lambda$, we found that the distributional density (which does not contribute to the effective mass) dominates the continuous density in the asymptotic time range, which is an unphysical effect. This same effect occurs for the negatively curved case ($\Lambda=0,\,K<0$). 

Furthermore, we raised the issue of whether the presence of this unphysical  distributional source could be detected by observations based on light rays crossing the timelike hypersurface made by the time evolution of the 3--sphere equator. By looking at radial null geodesics in the case $K=0=\Lambda$ and placing the observer at the symmetry centre $r=0$, we showed that the presence of the distributional source causes a discontinuos radial derivative of redshifts from observers beyond the equatorial hypersurface of $\bf{S}^3$. Hence, we proved that this type of distributional source would be detectable by observations, even if it does not contribute to the effective quasi--local mass. Finally, and for the purpose of comparison, we showed that this discontinuity of the redshifts does not occur in re-collapsing closed LTB models (for which there is no distributional source at the 3--sphere equator).  

\begin{acknowledgements}
SN acknowledges financial support from SEP--CONACYT postgraduate grants program.  RAS acknowledges financial support from SEP--CONACYT grant 239639.
\end{acknowledgements}

\appendix

\section{Calculation of limits}\label{sec:limit}
\subsection{$K>0$}
From \eqref{ellipclass} we have the following relations
$$\eta = \arccos\left( 1 - \alpha R\right),\;\;\; \sin \eta = \sqrt{1-\cos \eta} = \sqrt{\alpha R}\sqrt{2-\alpha R},$$
$$t-t_{bb} = \frac{1}{\beta}\left\lbrace\arccos\left( 1 - \alpha R\right)-\sqrt{\alpha R}\sqrt{2-\alpha R}\right\rbrace,$$
where $\alpha =\frac{K}{M}$ and $\beta=\frac{K^\frac{3}{2}}{M}=\alpha K^\frac{1}{2}$.

Derivating respect to $r$ and isolating $R'$ we obtain an expression which involves gradients which vanish at $\mathcal{S}$, so $R'$ vanishes also along the hypersurface.\\

Derivating once again and substituing
$$R''=\frac{1}{4}\frac{15M*D(F*B+G)+H+I+J}{K^4 RM(KR-2M)}$$
where
\begin{eqnarray*}
F&=&\left(K'^2M-\frac{4}{5}K'M'K-\frac{2}{5}K''KM+\frac{4}{15}M''K^2\right)(KR-2M),\\
G&=&\frac{4}{15}t_{bb}''\left(2MK^\frac{7}{2}-RK^\frac{9}{2}\right),\\
H&=&2K^4MR^3K''+8K^4MR^2M''-16K^3M^2R^2K''-16K^3M^2RM''+24K^2M^3RK'',\\
I&=&4K^4MR^2K'R'-3K^3MR^3K'^2+4K^4M^2R'^2-8K^4MRM'R'+4K^4R^2M'^2,\\
J&=&48K^2M^2RK'M'-60KM^3RK'^2-32K^3MR^2K'M'+40K^2M^2R^2K'^2.
\end{eqnarray*}
Note that at $\mathcal{S}$, $I$ and $J$ vanish. As not all functions vanish at $\mathcal{S}$, $R''$ is not necessarily of the form $0/0$. In general, in radial profiles $ KR\neq 2M$ so $R''$ is finite. From our choice of free functions $H$ doesn't vanish at $\mathcal{S}$.

We now analyze the term $K'/(2-2K)$, as the numerator and denominator are zero at the hypersurface, using the choice of free functions previously used we obtain that there is no limit at $\mathcal{S}$, so the term is singular. In the general case, L'H\^{o}pital's rule gives 
\begin{equation}\label{formacero}
\lim_{r\to\frac{\pi}{2}}\frac{K'}{2-2K}=\lim_{r\to\frac{\pi}{2}}\frac{K''}{K'}
\end{equation}
which necessarily gives a $0/0$ form, $\infty/0$ form or no limit as $K'$ is $0$ at the hypersurface.

The term 
$$\frac{\dot{R}'}{\sqrt{1-K}}=\frac{1}{2}\frac{\frac{2M'}{R}-\frac{2MR'}{R^2}-K'}{\sqrt{1-K}\sqrt{\frac{2M}{R}-K}}$$
clearly is of the form $0/0$ at $\mathcal{S}$. As $R', M', K'$ and $\sqrt{1-K}$ have zeros of the same order, this limit is well defined.



\end{document}